# Experimental observation of spin to charge current conversion at non-magnetic metal/Bi$_2$O$_3$ interfaces

Shutaro Karube[1,2], Kouta Kondou[2], and YoshiChika Otani[1,2]

[1] *Institute for Solid State Physics, University of Tokyo, Kashiwa 277-8581, Japan*

[2] *Center of Emergent Matter Science, RIKEN, 2-1 Hirosawa, Wako 351-0198, Japan*

We here demonstrate the interfacial spin to charge current conversion by means of spin pumping from a ferromagnetic Permalloy (Py: Ni$_{80}$Fe$_{20}$) to a Cu/Bi$_2$O$_3$ interface. A clear signature of the spin to charge current conversion was observed in voltage spectrum of a Py/Cu/Bi$_2$O$_3$ trilayer film whereas no signature in a Py/Cu and Py/Bi$_2$O$_3$ bilayer films. We also found that the conversion coefficient strongly depended on Cu thickness, reflecting the thickness dependent momentum relaxation time in Cu layer.

Conversion between spin and charge current is one of the most important phenomena in spintronics, which can be applied as a spin current source for domain wall displacement and magnetization switching.[1,2] The most common mechanism for spin current generation without using ferromagnets is spin Hall effects (SHEs), which occur in heavy metals such as Ta and Pt due to large spin-orbit coupling (SOC).[2,3] On the other hand, recently Rashba SOC has also been employed for the conversion. Rojas Sánchez *et al.* demonstrated that spin to charge current conversion took place at Bi(111)/Ag interface possessing a large Rashba SOC.[4] This phenomenon is known as the inverse Rashba-Edelstein effect (IREE),[5] and is also considered as a novel mechanism for spin conversion. Edelstein theoretically showed that non-equilibrium magnetization appears when charge current is flowing in two dimensional electrons gas.[6] Therefore, when the spin current polarized along $+y$ direction is injected vertically into the Rashba interface, the cross section of spin-split dispersion curve shown in Fig. 1(a) is shifted along $+k_x$ or $-k_x$ axis depending on the spin polarization (up or down), due to spin-momentum locking. This means that the charge current is generated by injection of the spin current, i.e. the spin to charge current conversion.

Following the report of Rojas Sánchez *et al.*, a few experimental papers related to IREE at Ag/Bi and Ag/Sb interfaces have been reported so far.[7,8,9] If an electric field can be applied efficiently at the Rashba interface, the spin to charge current conversion coefficient might be controlled electrically. However, it is challenging to apply the external electric field to the metallic interface. Therefore, it may be a better option to utilize a metal/insulator interface for applying an electric field. Previously we have reported that the enhanced spin relaxation took place at the Ag/$Bi_2O_3$ interface in the spin transport measurement, and assumed that the relaxation process was affected by Rashba SOC at the interface as a plausible scenario.[10] This result motivated us to perform the present study of the spin to charge current conversion using the metal/insulator interface. In this letter, we demonstrate that the spin to charge current conversion can be generated at a metal/insulator, $Cu/Bi_2O_3$ interface by means of spin pumping method.

Firstly, we describe our sample fabrication. The $Ni_{80}Fe_{20}$ (Py)/Cu/$Bi_2O_3$ trilayer, Py/Cu bilayer and Py/$Bi_2O_3$ bilayer were prepared on a Si substrate with 300 nm of $SiO_2$ layer by lift-off using suspended resist mask and e-beam evaporation in a high vacuum of about $10^{-7}$ Pa. The mask consisting of primer (hexamethyl-disilazane) and AZ1500 resist layers was patterned into a 5 μm wide 200 μm long strip by maskless photolithography. After the deposition, lift-off process was performed by soaking the samples into acetone for a few hours. We also fabricated a Ti(5 nm)/Au(200 nm) waveguide neighboring the sample wire for inducing ferromagnetic resonance (FMR) in Py during the spin pumping measurements as shown in Fig. 1(b). $Al_2O_3$ dielectric layer was deposited to prevent shunting between the strip and the waveguide. During the spin pumping measurement, the voltage across the strip was measured by sweeping the in-plane magnetic field applied perpendicular to it with passing a 9 GHz rf current in the waveguide.

Here we show the experimental results of spin to charge current conversion at the $Cu/Bi_2O_3$ interface. All the measurements were performed at room temperature (RT). We prepared three different multilayer strips, Py(5nm)/Cu(10nm)/$Bi_2O_3$(100nm), Py(5nm)/Cu(10nm), and Py(5nm)/$Bi_2O_3$(100nm) to elucidate an importance of a $Cu/Bi_2O_3$ interface. As shown in Fig. 1(c), by means of spin pumping, we injected spin currents into the $Cu/Bi_2O_3$ interface or Cu. If the spin to charge current conversion takes place at the Cu/Bi$_2$O$_3$ interface, the peaks should be observed in the rectified voltage spectra as a function of magnetic field. As expected, the clear peak corresponding to the spin to charge current conversion was observed in the voltage spectrum of a Py/Cu/Bi$_2$O$_3$ trilayer strip whereas no peaks in the case of a Py/Cu and Py/Bi$_2$O$_3$ bilayer strips, at the resonant field with a rf current of 9 GHz and 23 mW as shown in Fig. 2(a), (b) and (c), respectively. This result supports an importance of Cu/Bi$_2$O$_3$ interface for the conversion. Considering that Bi$_2$O$_3$ is an insulator and Cu does not contribute to the conversion, the above experimental facts suggest that this result comes from an interfacial phenomenon. We also investigated a half width at half maximum (HWHM) $\Delta H$ of a FMR spectrum of Py as a function of frequency for the Py/Cu/Bi$_2$O$_3$, Py/Cu and Py/Bi$_2$O$_3$ films by using a vector network analyzer as shown in Fig. 1(a), (b) and (c). The Py/Cu/Bi$_2$O$_3$ film shows steeper linearity to the rf frequency $f$ than the Py/Cu, as shown in Fig. 2(d). This means that the SOC of Cu/Bi$_2$O$_3$ interface enhances damping of FMR in Py, i.e. the spin current is injected into the interface. From the slope, we can estimate a damping constant $\delta$ for Py using the following equation,[11] $\mu_0\Delta H(f) = \mu_0\Delta H(0) + (2\pi\delta/\gamma_e)f$, where $\gamma_e = g\mu_B/\hbar$ and $\mu_0\Delta H(0)$ are the gyromagnetic ratio of electrons and the offset of HWHM, respectively. Here $g$, $\mu_B$ and $\hbar$ are the Lande's g-factor of 2.1,[12] the Bohr magnetron and the reduced Planck constant, respectively. We then obtain the damping constants $\delta = 0.0140$, $0.0095$ and $0.0173$ for Py/Cu/Bi$_2$O$_3$, Py/Cu and Py/Bi$_2$O$_3$, respectively. The value of $\delta$ for Py/Bi$_2$O$_3$ is larger than those of the other cases because the damping constant of Py is enhanced possibly due to strong SOC of Bismuth atoms. The difference in the damping constant gives the spin injection efficiency known as spin mixing conductance,[13]

$$g_{\text{eff}}^{\uparrow\downarrow} = \frac{4\pi M_S t_F}{g\mu_B}\left(\delta_{F/N/O} - \delta_{F/N}\right), \tag{1}$$

where $\mu_0$, $M_S$, $t_F$, $\delta_{F/N/O}$, $\delta_{F/N}$ are vacuum permeability, saturation magnetization, thickness of Py, damping constant for Py/Cu/Bi$_2$O$_3$, and for Py/Cu, respectively. Here $\mu_0 M_S$ is determined by fitting the Kittel formula to the data as shown in Fig. 2(e). The injected spin current density $J_S$ is given by Eq. (2),[14]

$$J_S = \frac{2e}{\hbar} \times \frac{\hbar g_{\text{eff}}^{\uparrow\downarrow} \gamma_e^2 (\mu_0 h_{\text{rf}})^2 \left[\mu_0 M_S \gamma_e + \sqrt{(\mu_0 M_S)^2 \gamma_e^2 + 4\omega^2}\right]}{8\pi \delta_{F/N/O}^2 [(\mu_0 M_S)^2 \gamma_e^2 + 4\omega^2]} \times \exp\left(-\frac{t_N}{\lambda_{\text{sf}}^N}\right) \tag{2}$$

where $h_{\text{rf}}$, $\omega = 2\pi f$ are the applied rf field and the angular frequency, respectively. The injected spin current at the Py/Cu interface propagates and exponentially decays in the Cu layer. The spin diffusion length $\lambda_{\text{sf}}^N$ of Cu is estimated by assuming the relation $\lambda_{\text{sf}}^N \propto 1/\rho_{\text{Cu}}$ with the resistivity of Cu; $\rho_{\text{Cu}}$. The value of $\lambda_{\text{sf}}^N$ is calculated from the ratio of $\rho_{\text{Cu}}$ in the present study to the reported $\rho_{\text{Cu}} \sim 3$ μΩcm with considering $\lambda_{\text{sf}}^N \sim 300$ nm at RT.[15] This term does not almost affect the original $J_S$ injected at the Py/Cu interface. As soon as it reaches the Cu/Bi$_2$O$_3$ interface, the spin current is converted to the charge current. The resulting charge current density $j_C$ flowing in the two dimensional interface is expressed as $j_C = I_C/w = V/wR_{\text{tot}}$, where $w$, $I_C$, $V$, $R_{\text{tot}}$ are the width of theسample wire, the charge current, the detected voltage, and the total resistance of the wire, respectively. From these densities, we can calculate the conversion coefficient $\lambda_{\text{IREE}} = j_C/J_S$. Here the units of $j_C$ and $J_S$ are A/m and A/m$^2$, respectively. Therefore $\lambda_{\text{IREE}}$ has a unit of length. Fig. 3(a) shows the Cu thickness dependence of $\lambda_{\text{IREE}}$, including the values of the Py/Bi$_2$O$_3$ bilayer in $t_{\text{Cu}} = 0$ nm. The values of $\delta_{F/N/O}$ and $g_{\text{eff}}^{\uparrow\downarrow}$ which are utilized for the calculation of the $\lambda_\text{IREE}$ are independent of the Cu layer thickness. And the magnitude of the $g_\text{eff}^{\uparrow\downarrow}$ in the Py/Cu/Bi$_2$O$_3$ trilayer is $(1.1 \pm 0.2) \times 10^{19}$ m$^{-2}$ which is comparable to that of metal/metal Rashba interface such as Bi(111)/Ag[4,9], implying that the spin injection efficiency due to spin pumping is almost same. On the other hand, $\lambda_\text{IREE}$ strongly depends on Cu thickness as shown Fig. 3 (a). The amplitude of $\lambda_\text{IREE}$ increases with increasing the Cu thickness and is saturated at 0.6 nm above $t_\text{Cu}$ = 15 nm. The sign of the $\lambda_\text{IREE}$ is opposite compared with previously reported values in Ag/Bi interface,[4,7] and the trend is the same for the spin directions of Fermi contour with Rashba splitting between Cu/Bi[16,17] and Ag/Bi interface.[18,19]

Surprisingly the saturated value of $\lambda_\text{IREE}$ is twice larger than the reported value ($\lambda_\text{IREE}^\text{Bi/Ag}$ = 0.3 nm) in epitaxial grown Bi(111)/Ag interface.[4] In case of metal/metal interface such as Bi/Ag, a part of injected spin current might be absorbed not only at interface but also in the adjacent bulk layer. Therefore, reported $\lambda_\text{IREE}$ in metal/metal interface is lower limit value of $\lambda_\text{IREE}$. While in the case of metal/oxide interface, the injected spin current is converted to the charge current at the interface. Thus, the metal/oxide interface is suitable sample structure to investigate the interfacial spin to charge current conversion phenomena.

Finally, we discuss the origin of Cu thickness dependent $\lambda_\text{IREE}$ expressed by using Rashba parameter $\alpha_\text{R}$ and momentum relaxation time $\tau_e^\text{int}$ at the interface,[5]

$$\lambda_\text{IREE} = \frac{\alpha_\text{R} \tau_e^\text{int}}{\hbar}, \qquad (3)$$

So far, the reported values of $\tau_e^\text{int}$ in two dimensional surface or interface states of either topological insulators or quantum wells with Rashba splitting have been in the range of several picoseconds, all of which were determined by means of pump probe spectroscopy (PPS)[20] and angle-resolved photoemission spectroscopy (ARPES).[21] However, the momentum relaxation time of Cu, $\tau_e^{Cu}$ of our sample was determined to be 4 - 9 femtoseconds. This indicates that the $\tau_e^{int}$ is governed by the momentum relaxation in the bulk Cu layer. Therefore the estimated value of $\lambda_{IREE}$ strongly depends on the $\tau_e^{Cu}$ whose value was calculated from a resistivity $\rho_{Cu}$ of Cu, as shown in Figure 3 (b). Figure 3(c) shows the Cu thickness dependence of $\alpha_R$ estimated by using $\tau_e^{Cu}$ instead of $\tau_e^{int}$. The $\alpha_R$ was estimated to be -(0.46 $\pm$ 0.06) eV·Å, which is about half of that for the Cu(111)/Bi interface.[22] This difference may be caused by following reasons: Bismuth and Oxygen atoms are arranged randomly on the Cu layer because deposited $Bi_2O_3$ layer is amorphous, and therefore the atomic contact area of Bi/Cu interface is smaller than that at the Cu(111)/Bi interface; the above described situation possibly reduces the magnitude of SOC at the $Cu/Bi_2O_3$ interface compared with the Cu/Bi interface; the existence of Oxygen atoms is also considered to make another contribution. Since the oxygen atom has strong electronegativity, a significant part of charge density near Bi atoms will be attracted towards the oxygen atom, resulting in the reduced Rashba effect because of the decreased charge density for producing an interfacial electric field.[16]

In summary, we have demonstrated the spin to charge current conversion in $Py/Cu/Bi_2O_3$ trilayer film by means of spin pumping method. Comparing with the spectra of the Py/Cu, $Py/Bi_2O_3$ and $Py/Cu/Bi_2O_3$ films, we confirmed that the results are originated from the $Cu/Bi_2O_3$ interface. Furthermore we found that the $\lambda_{IREE}$ is strongly influenced by the momentum relaxation in bulk Cu layer. We expect that this non-magnetic metal/$Bi_2O_3$ type interface pave a way for controlling the conversion between spin and charge current by electric field effect, which could be beneficial for domain wall motion and the magnetization switching.

**Acknowledgments**

We thank Y. Fukuma for his constructive suggestions. This work was supported by Grant-in-Aid for Scientific Research on Innovative Area, "Nano Spin Conversion Science" (Grant No. 26103002) and Japan Society for the Promotion of Science through Program for Leading Graduate Schools (MERIT).

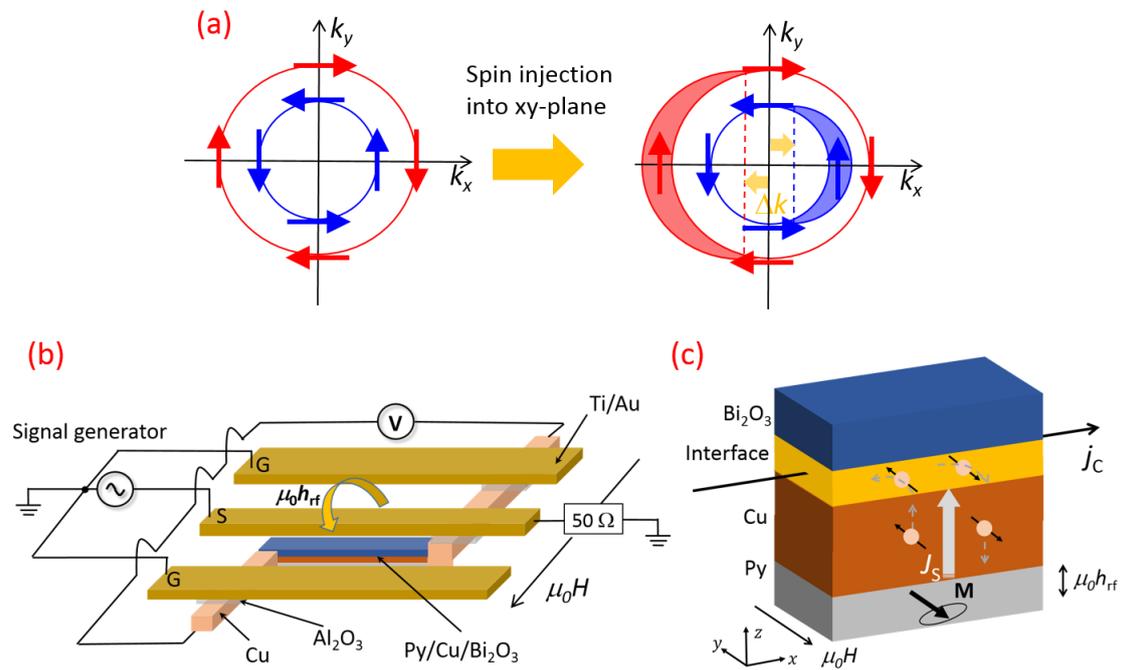

Figure 1. (a) Fermi contours with Rashba splitting and non-equilibrium spin density resulted as a shift of Fermi contours by spin injection. (b) Experimental setup for the spin pumping measurement. (c) Schematic image of spin to charge current conversion at the Cu/$Bi_2O_3$ interface. Spin currents are generated by spin pumping from a Py layer in ferromagnetic resonance to a Cu layer and reach a Cu/$Bi_2O_3$ interface. Then, the spin to charge current conversion takes place due to the spin-momentum locking at the Rashba interface.

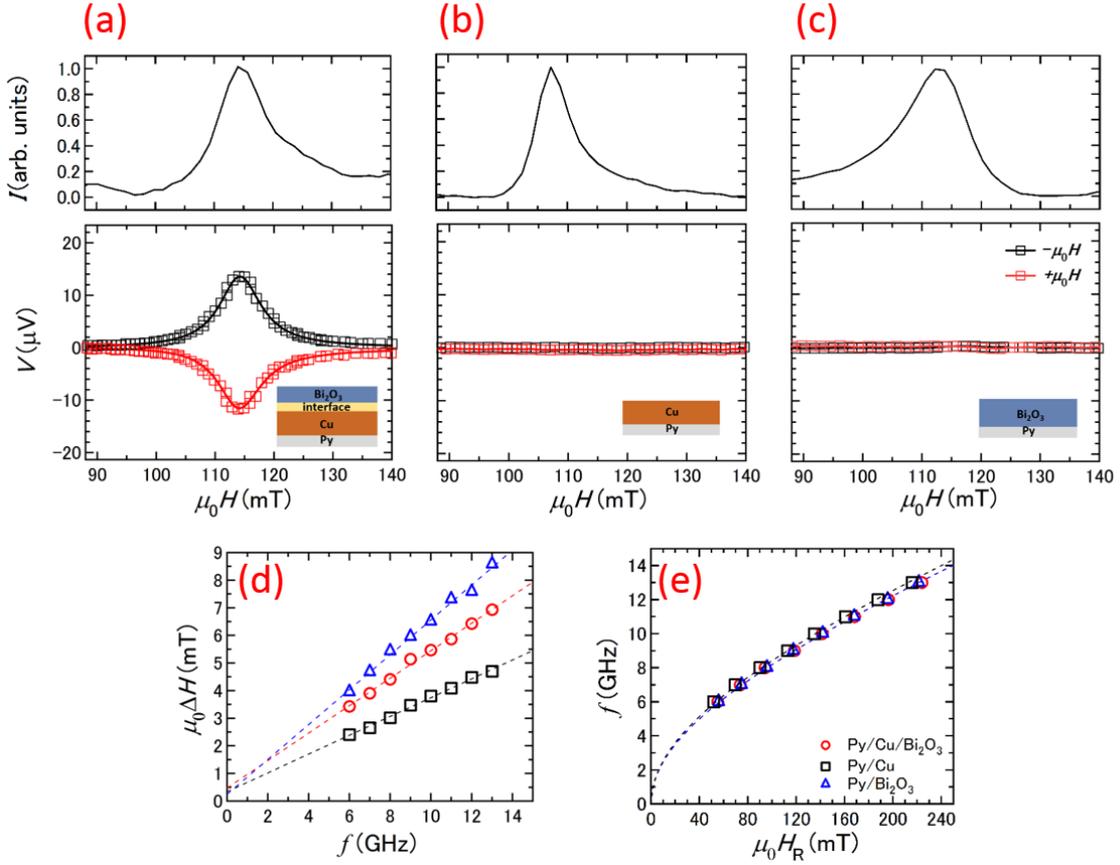

Figure 2. Comparison between Py/Cu/Bi$_2$O$_3$ trilayer, Py/Cu bilayer, and Py/Bi$_2$O$_3$ bilayer. Magnetic field dependence of the microwave absorption $I$ and the rectified voltages $V$ in (a) Py/Cu/Bi$_2$O$_3$ trilayer, (b) Py/Cu bilayer and (c) Py/Bi$_2$O$_3$ bilayer. Plot and solid lines (red and black) are the raw data and fitting curve, respectively. (d) Half width at half maximum $\mu_0 \Delta H$ of the spectrum in each case. (e) Applied frequency of rf current as a function of the resonant magnetic field, where the dashed lines mean fitting curve by Kittel formula.

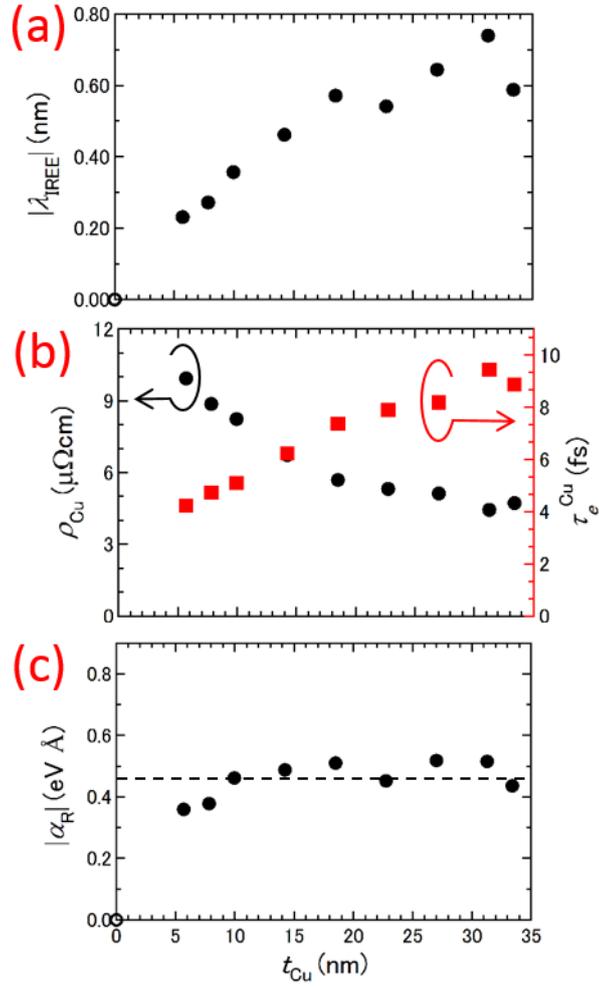

Figure 3. Cu thickness dependences of (a) the magnitudes of the conversion coefficient $\lambda_{\mathrm{IREE}}$, (b) the resistivity $\rho_{\mathrm{Cu}}$ and the momentum relaxation time $\tau_e^{\mathrm{Cu}}$ of Cu layer, and (c) the Rashba parameter $\alpha_{\mathrm{R}}$, respectively. In (b), the right red axis stands for $\tau_e^{\mathrm{Cu}}$. Here open circle at $t_{\mathrm{Cu}} = 0$ nm in (a) and (c) means data for Py/Bi$_2$O$_3$. Note that the signs of $\lambda_{\mathrm{IREE}}$ and $\alpha_{\mathrm{R}}$ are negative.